# Numerical Simulation of Iced Wing Using Separating Shear Layer Fixed Turbulence Models

Haoran Li,[1,*]  Yufei Zhang [1,†], Haixin Chen [1, ‡]
*(1. Tsinghua University, Beijing, 100084, People's Republic of China)*

**Aerodynamic prediction of glaze ice accretion on airfoils and wing is studied using the Reynolds-averaged Navier-Stokes method. Two separation fixed turbulence models are developed considering the nonequilibrium characteristics of turbulence. The key ad hoc fix is a term of the local ratio of turbulent production to dissipation, which is used to amplify the destruction term of the $\omega$-equation to increase the eddy viscosity in a separating shear layer of the fully turbulent region. A shear stress limiter is adopted to appropriately simulate the beginning process of the shear layer transition when the turbulence is under development. The proposed separation fixed terms can be easily implemented into current solvers. Two airfoils and a three-dimensional swept wing with ice accretions are numerically tested using the modified models. The results indicate that the separating shear layer fixes improve the ability of the models in predicting the stall behavior at large angles of attack. The simulated averaged flow field and turbulence intensity distribution are consistent with experimental data.**

## Nomenclature

$h_{ice}$      = ice height

$c$      = airfoil chord length

Re      = Reynolds number

$\rho$      = air density

$\Omega$      = magnitude of vorticity

---

[*] Ph. D student, School of Aerospace Engineering, email: lihr17@mails.tsinghua.edu.cn
[†] Associate professor, School of Aerospace Engineering, senior member AIAA, email: zhangyufei@tsinghua.edu.cn
[‡] Professor, School of Aerospace Engineering, associate fellow AIAA, email: chenhaixin@tsinghua.edu.cn

| | |
|---|---|
| $d$ | = distance to the closest wall |
| $S_{ij}$ | = strain rate tensor |
| $AOA$ | = angle of attack |
| $C_L$ | = lift coefficient |
| $C_D$ | = drag coefficient |
| $C_m$ | = pitching moment coefficient |
| $C_p$ | = pressure coefficient |
| $U_{rms}$ | = root mean square of velocity fluctuation |

## I. Introduction

The performance of aircraft in icing conditions suffers a great deterioration due to ice accretion on lifting surfaces without anti or deicing systems. The initial cost, cost of maintenance, and weight effect associated with ice protection systems make their application are practical only for the most critical components [1]. Ice accretion on the unprotected lifting surface appears as a protrusion that changes the streamline shape into a nonstreamline body and degrades the aerodynamic performance [2]. The effects of ice accretion on the aerodynamic performance of a wing are a complicated problem that affects the design and airworthiness certification of a civil transport aircraft.

The ice accretion effect on aerodynamic performance has been studied by experimental and numerical methods for many years. Earlier measurements for aerodynamic performance of iced airfoils can be traced back to the 1940s when some air accidents were first diagnosed as a result of aircraft icing [3]. A catalog of ice shapes under a wide range of conditions and their effects are formed though experiments. Among all the ice shapes, researchers have generally focused on the rime and glaze-type ice because they are the commonest. Moreover, glaze ice shapes should receive more attention because they can cause severe degradation of aerodynamic performance. The experimental work of Bragg et al. [4] and the corresponding computational research of Potapczuk [5] focused on a two-dimensional (2D) NACA0012 airfoil with simulated glaze ice accretion. They found a great loss in lift and excessive drag induced by glaze ice. Another typical glazed ice airfoil is the GLC305 airfoil with 944 ice shape, which has abundant experimental data. The GLC305 airfoil is a section of a business jet wing, and its glaze icing condition is selected from the Federal Aviation Administration's FAR Part 25-Appendix C Atmospheric Icing Conditions. Additional comprehensive flow field information such as the stagnation streamline location is provided by Broeren et al. [6] [7]. Most icing measurements for aerodynamic evaluating purposes have only addressed two-dimensional aircraft components. Bragg et al. [1] first investigated the aerodynamics performances about a wing with manually installed glaze ice. Khodadoust and Bragg et al. [8] extended three-dimensional (3D) wing pressure measurements to test the effect of the wing sweep and studied the iced performances. The model in their paper is one of the few public 3D geometries of the iced wing. Sufficient data with high-quality to evaluate the ice accretion on wings are currently unavailable in the public domain [9]. Since the high-fidelity ice accretion of a swept wing shows great nonuniformity along the span with the shape similar to "scallops" or "lobster tail", it remains unclear how much detail of the three-dimensionality is critical to the aerodynamics and can be appropriately simulated.

There have been several investigations for the 2D Reynolds-averaged Navier-Stokes (RANS) simulations of iced airfoils. The consensus is that the turbulence model is the most important factor to accurately predict the performance of an iced airfoil. Potapczuk [5] studied the aerodynamic performance of glaze ice on the leading edge of an NACA0012 airfoil with the eddy viscosity Baldwin-Lomax algebraic two-layer model. The predicted lift, drag, and moment coefficients are consistent under angles of attack below stall. For angles of attack after stall, time-averaged results show a significant deviation from the experiment. Dompierre et al. [10] reported computation results of iced airfoil using the finite-volume Galerkin method and $k - \varepsilon$ turbulence model, but no experimental data were available for comparison. The one-equation Spalart–Allmaras (SA) model [11] and two-equation $k - \omega$-based shear stress transport (SST) model [12] and are most popular in predicting the turbulence around an aircraft. Marongiu et al. [13] tested three RANS solvers (ZEN, FLUENT, and CFL3D) with SA and SST models in characterizing an iced airfoil with a double horn at the leading edge. They reported that the disagreement between these two models increases as the angle of attack augments. In addition, unsatisfied prediction results of the separated region on the upper surface have been observed at AOA=6 degrees. Costes and Moens [14] studied eddy-viscosity models SA and SST as well as Reynolds stress model models DRSM (Differential Reynolds Stress Model) and EARSM (Explicit Algebraic Reynolds Stress Model) in predicting the flow field of iced airfoils. They found that SA model is the only eddy-viscosity model converging efficiently towards steady-state, while DRSM provided natural unsteady solutions which have to be time-averaged.

Most of the RANS models, along with the implicit equilibrium approximations, can hardly be expected to provide an accurate description for separated flows. Rotta [15] concludes from analysis of experimental data that when a turbulent boundary layer is perturbed from its equilibrium state, a new equilibrium state is not attained for at least 10 boundary layer thickness downstream of the perturbation. In other words, separated flows are very much out of equilibrium. Coefficients of the described RANS models are calibrated for homogenous shear flow where the production to dissipation rate of turbulence kinetic energy ($P_k/\varepsilon$) is between 1.4 and 1.8, and the turbulence is in the equilibrium state [16]. For example, in SST model, the coefficients for ω-production and destruction terms are calibrated by this way. Therefore, the turbulence shear stress is under-predicted in a separating shear layer. Attempts have been made to remedy the problem of poor non-equilibrium characteristic. Shang and Hankey [17] introduced the notion of a relaxation length to reduce the Reynolds stress from equilibrium value predicted by the Cebeci-Smith model. Li et al. [18] developed a separating shear layer fixed (SPF) $k - \overline{v^2} - \omega$ model to simulate the iced airfoils, especially the stall behavior. The three-equation model focused on the transition process and the nonequilibrium characteristics of turbulence in separation regions. The nonequilibrium effect in a separating shear layer is that the turbulence production $P_k$ tends to be significantly larger than the dissipation ε [19], and often as high as 3-4 [20]. With the improved performance in predicting the shear layer, the results of the previous model are consistent with measurements for many types of iced airfoil. However, the SPF $k - \overline{v^2} - \omega$ model lacks validation for the simulation in three-dimensional cases, and the model is not widely applied in engineering applications. A more robust and widely used turbulence model that can predict the stall behavior of an iced wing is expected.

In this paper, we apply the separation fixes similar to the SPF $k - \overline{v^2} - \omega$ model to the widely used Wilcox 2006 $k - \omega$ model. The $k - \omega$ model was first independently proposed by Kolmogorov [21] and Saffman [22]. Wilcox [23] [24] [25] has continually built and improved this two-equation model, and demonstrated its accuracy for a wide range of turbulent flows. With the proposed modification terms, this paper introduces a separating shear layer fixed $k - \omega$ model. A flat plate case is selected to test the performance of turbulent models in predicting the log-layer. Two iced airfoils (NACA0012 and GLC305 airfoils) with glaze ice are numerically studied to show the performance of the new models in predicting the separated flow field. Research on the aerodynamics prediction of iced swept wing focuses on the models' ability for three-dimensional flows. Comparisons among three turbulence models (baseline $k - \omega$, SPF $k - \overline{v^2} - \omega$ and SPF $k - \omega$ models) are performed in this paper.

## II. Modifications of Turbulence Models

### A. Separating Shear layer fixes of SPF $k - \overline{v^2} - \omega$ model

The SPF $k - \overline{v^2} - \omega$ model transport equations are solved for the velocity fluctuation energy $k$ (the sum of fully turbulent and pretransition velocity fluctuations), $\overline{v^2}$ (fully turbulent fluctuations) and scale-determining variable ($\omega$) defined here as specific dissipation rate. The transport equations are listed as following where the $R_{BP}$ and $R_{NAT}$ terms represent for the bypass and natural transition.

$$\frac{\partial k}{\partial t} + u_j \frac{\partial k}{\partial x_j} = \frac{1}{\rho} P_k - min(\omega k, \omega \overline{v^2}) - \frac{1}{\rho} D_k + \frac{1}{\rho} \frac{\partial}{\partial x_j}\left[\left(\mu + \frac{\rho \alpha_T}{\sigma_k}\right) \frac{\partial k}{\partial x_j}\right] \quad (1)$$

$$\frac{\partial \overline{v^2}}{\partial t} + u_j \frac{\partial \overline{v^2}}{\partial x_j} = \frac{1}{\rho} P_{\overline{v^2}} + R_{BP} + R_{NAT} - \omega \overline{v^2} - \frac{1}{\rho} D_{\overline{v^2}} + \frac{1}{\rho} \frac{\partial}{\partial x_j}\left[\left(\mu + \frac{\rho \alpha_T}{\sigma_k}\right) \frac{\partial \overline{v^2}}{\partial x_j}\right] \quad (2)$$

$$\begin{aligned}\frac{\partial \omega}{\partial t} + u_j \frac{\partial \omega}{\partial x_j} &= \frac{1}{\rho} P_\omega + \left(\frac{C_{\omega R}}{f_W} - 1\right)\frac{\omega}{\overline{v^2}}(R_{BP} + R_{NAT}) - f_{NE} C_{\omega 2} \omega^2 f_W^2 \\ &+ 2\beta^*(1 - F_1^*)\sigma_{\omega 2} \frac{1}{\omega} \frac{\partial k}{\partial x_j} \frac{\partial \omega}{\partial x_j} + \frac{1}{\rho} \frac{\partial}{\partial x_j}\left[\left(\mu + \frac{\rho \alpha_T}{\sigma_\omega}\right) \frac{\partial \omega}{\partial x_j}\right]\end{aligned} \quad (3)$$

The separation ad hoc modification term $f_{NE}$ in Eq. (3) considers the nonequilibrium behavior of the turbulence. The model coefficients of baseline $k - \overline{v^2} - \omega$ is calibrated for equilibrium turbulence where the turbulence production-to-dissipation ratio is $P_{\overline{v^2}}/\varepsilon < 1.5$. Such a ratio exists in the boundary layer, wakes, jets, etc. [19] [26]. However, the ratio should be greatly larger than 1.5 in a separated shear layer, which indicates a fast growth of the turbulence provoked by the rapid breakdown of the turbulence [27]. $f_{NE}$ is multiplied to the destruction term of $\omega$ to help produce more shear stress in the fully turbulent region. In the previous work [14] [18], switch function $\Gamma_{SSL}$ is used to locate the separating shear layer region where $P_{\overline{v^2}}/\varepsilon > 2.5$, which means that the nonequilibrium modification term $f_{NE}$ is turned off elsewhere. The region with high $Re_\Omega$ in a shear layer must be the fully turbulent region with high Reynolds stress, and it can be used to determine the magnitude of the modification. The maximum value of $f_{NE}$ is chosen as 3.3 merely to keep unbounded and the fix is not believed to be particularly sensitive to this number.

$$f_{NE} = min(max(300 Re_\Omega \Gamma_{SSL}, 1), 3.3), \quad Re_\Omega = \frac{d^2 \Omega}{\nu} \quad (4)$$

$$\Gamma_{SSL} = \frac{1}{1 + e^{-10(\frac{P_{\overline{v^2}}}{\varepsilon} - C_{SSL})}}, \quad \frac{P_{\overline{v^2}}}{\varepsilon} = \frac{\mu_{T,s} S^2}{\rho \overline{v^2} \omega}, \quad C_{SSL} = 2.5 \quad (5)$$

The separating shear layer of an iced airfoil experiences a transition process from laminar state to turbulent flow until reattachment [28] [29]. Eq. (6) is a form of shear stress limiter included in the SPF $k - \overline{v^2} - \omega$ model, which acts on the beginning region of separating shear layer just leaving the ice tip where the turbulence is under development.

$a_1$ and $a_2$ are proportional coefficients used particularly for the boundary and separating shear layer, where $a_1$ is set as 0.31, and $a_2$ is 0.23. The switch between the two coefficients is determined by the function $\Gamma_{SSL}$.

$$\mu_{T,s} = min\left[\rho f_v f_{INT} C_\mu \sqrt{\overline{v_s^2}} \lambda_{eff}, min\left(a_1, \frac{a_2}{\Gamma_{SSL}}\right) \frac{\rho \overline{v^2}}{\Omega F_2}\right] \tag{6}$$

Some other modifications that specific to the transport equations of SPF $k - \overline{v^2} - \omega$ model, like changing the characteristic length scale and constant model parameters, are not presented currently. Detailed explanations can be found in [18].

## B. Two-equation SPF $k - \omega$ model

Our previous study shows that the SPF $k - \overline{v^2} - \omega$ model includes some physical based modeling features to predict separated flows, especially nonequilibrium modifications. However, the three-equation transitional model is not widely used in engineering applications. Thus, we begin to explore whether the modifications can be adapted to a widely used model. The fully turbulent Wilcox 2006 $k - \omega$ model [24] has a similar framework to the $k - \overline{v^2} - \omega$ model. The $k - \overline{v^2} - \omega$ model yields $\overline{v^2} \approx k$ in the full turbulent area, where the model could be effectively reduced to the form of the $k - \omega$ transport equations. Then, it is believed that the separated modification of the SPF $k - \overline{v^2} - \omega$ model can also be adapted to the $k - \omega$ model. The Wilcox 2006 $k - \omega$ model is used as the baseline model. The modified Wilcox 2006 $k - \omega$ model in this paper is referred to as the separating shear layer fixed $k - \omega$ model and be abbreviated as the SPF $k - \omega$ model.

The two-equation SPF $k - \omega$ model is as follows:

$$\frac{\partial k}{\partial t} + u_j \frac{\partial k}{\partial x_j} = \frac{1}{\rho} P_k - \beta^* \omega k + \frac{1}{\rho} \frac{\partial}{\partial x_j}\left[\left(\mu + \frac{\rho k}{\sigma_k \omega}\right) \frac{\partial k}{\partial x_j}\right] \tag{7}$$

$$\frac{\partial \omega}{\partial t} + u_j \frac{\partial \omega}{\partial x_j} = \frac{1}{\rho} P_\omega - f_{NE} \beta \omega^2 + \frac{\sigma_d}{\omega} \frac{\partial k}{\partial x_j} \frac{\partial \omega}{\partial x_j} + \frac{1}{\rho} \frac{\partial}{\partial x_j}\left[\left(\mu + \frac{\rho k}{\sigma_\omega \omega}\right) \frac{\partial \omega}{\partial x_j}\right] \tag{8}$$

The nonequilibrium modification term $f_{NE}$ in Eq. (9) has an identical form to that of the SPF $k - \overline{v^2} - \omega$ model, but there are differences in the upbound of $f_{NE}$ and threshold of $P_k/\varepsilon$. The upbound is enlarged from 3.3 in the SPF $k - \overline{v^2} - \omega$ model to 6.0 in the SPF $k - \omega$ model. Threshold value $C_{SSL}$ is decreased from 2.5 to 1.5. Such adjustment is calibrated by the separated flow field of an iced airfoil - Test Case B in the following part. The calibration determined by an iced airfoil case can be adopted to other iced airfoil cases mainly because the nonequilibrium flow inside the separating shear layer has a same characteristic: $P_k$ tends to be significantly larger than the dissipation $\varepsilon$, and often as high as 3-4. Although the modification functions seem different, the goal is to adjust the same level of the nonequilibrium characteristic. The modification term does not affect the simulation of the attached flows, i.e., the calculation of the boundary layer will be identical to the original $k-\omega$ model because $P_k/\varepsilon$ is supposed to be lower than 1.5, and $\Gamma_{SSL}$ is close to 0 in such flows.

$$f_{NE} = min(max(300Re_\Omega \Gamma_{SSL}, 1), 6.0), \quad Re_\Omega = \frac{d^2 \Omega}{\nu} \tag{9}$$

$$\Gamma_{SSL} = \frac{1}{1 + e^{-10(\frac{P_k}{\varepsilon} - C_{SSL})}}, \quad \frac{P_k}{\varepsilon} = \frac{\mu_T \Omega^2}{\beta^* \rho k \omega}, \quad C_{SSL} = 1.5 \tag{10}$$

The second key modification to the SPF $k-\omega$ model refers to the shear stress limiter. The turbulent viscosity inside the initial position of the separating shear layer can be more accurately simulated by limiting the magnitude of Reynolds shear stress. Because the flow there is in a transition state, the turbulent intensity at the beginning of the transition process is small. Based on the successful experience of the SPF $k-\overline{v^2}-\omega$ model, we introduce an ad hoc fix to Wilcox's shear stress limiter for $\widetilde{\omega}$ as Eq. (11). $\mu_T = \rho k/\widetilde{\omega}$ in Wilcox 2006 $k-\omega$ model [24]. The shear stress limiter can drive turbulent shear stress towards the relational expression of Bradshaw et al. [30]. In a shear layer, $2\bar{S}_{ij}\bar{S}_{ij} \approx (\partial u/\partial y)^2$, and Eq. (11) tells us the form of Eq. (12). When $C_{lim} = 1$, the coefficient $C_{lim}^{-1}\sqrt{\beta^*} = 0.30$ matches Bradshaw's constant. In the SPF $k-\omega$ model to calculate the iced airfoil, $C_{lim} = 0.875$ is retained as the original $k-\omega$ model, and $C_{lim2}$ is set as 1.2.

$$\widetilde{\omega} = max\left[\omega, max(C_{lim}, C_{lim2}\Gamma_{SSL})\sqrt{\frac{2\bar{S}_{ij}\bar{S}_{ij}}{\beta^*}}\right] \quad C_{lim} = 0.875 \quad C_{lim2} = 1.2 \tag{11}$$

$$\rho\tau_{xy} = \mu_T \frac{\partial u}{\partial y} = min\left[\frac{\rho k}{\omega}\frac{\partial u}{\partial y}, min\left(C_{lim}^{-1}, \frac{C_{lim2}^{-1}}{\Gamma_{SSL}}\right)\sqrt{\beta^*}\rho k\right] \tag{12}$$

### III. Numerical Method

Aerodynamic analysis in this paper adopts the RANS solver CFL3D [31] with structured grids. The governing equations, which are the thin-layer approximations to the three-dimensional time-dependent compressible Navier-Stokes (N-S) equations are solved. Both N-S and turbulence model equations are nondimensionalized. The spatial discretization is based on a cell-centered finite volume formulation, with third-order upwind-biased differencing. The code is advanced in time with an implicit approximate-factorization method. Several convergence acceleration options are available including multigrid and mesh sequencing. All turbulence field-equation models are solved uncoupled from Navier-Stokes equations under essentially identical fashion. For eddy viscosity models, the process of solving the turbulent Navier-Stokes equations are identical to the laminar equations with the exception that $\mu$ is replaced by $\mu + \mu_T$, where $\mu_T$ is the eddy viscosity value obtained by whatever turbulence model is used. All of the field equation models except for Wilcox 2006 $k - \omega$ make use of the distance to the nearest wall. The following cases show good performance of residual convergence.

## IV. Test Cases

**A. Flat Plate Log-Layer**

In the boundary log-layer, experiments show that the production is approximately in equilibrium with the dissipation $P_k \approx \varepsilon$, and the ratio of $\tau_{xy}$ to $k$ is about 0.3. The described modifications should not affect these laws. Therefore, a flat plate case is selected to test the performance of turbulent models in predicting the boundary layer. Subsonic flow past a semi-infinite flat plate is modeled at Reynolds number $5 \times 10^6$ and Mach number 0.2. Inflow is set by specifying total pressure and total temperature where $p_{total}/p_{inf} = 1.02828$ and $T_{total}/T_{inf} = 1.008$. The wall starts at $x=0$. The velocity profiles at two positions $x = 0.3$ and $x = 0.8$ are extracted. The $k - \omega$ (in this paper, the $k - \omega$ all refer to Wilcox 2006 $k - \omega$ version), $k - \overline{v^2} - \omega$ and modified SPF $k - \omega$ and SPF $k - \overline{v^2} - \omega$ models yield acceptable results compared with the theoretical data [32]. The following figures illustrate the case.

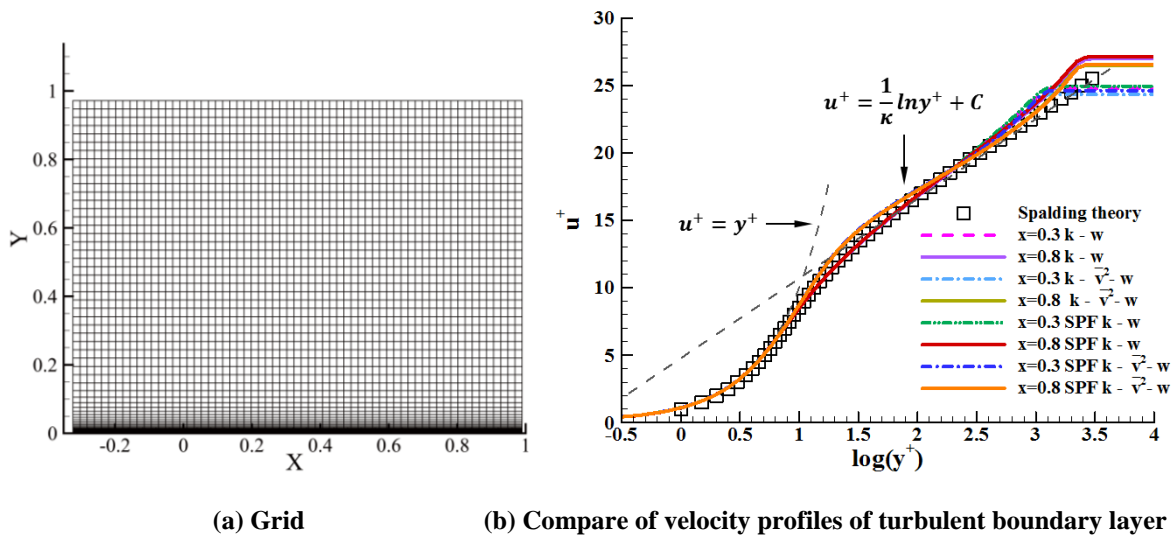

(a) Grid      (b) Compare of velocity profiles of turbulent boundary layer

**Fig. 1. Computational domain and comparison of velocity profiles**

## B. Periodic hill

Flow pass through a periodic hill is investigated to validate the improved RANS methods. The prediction of separating shear layer and reattachment point can be a challenge for turbulent models [33]. Fig. 2 shows the computational domain and boundary conditions. The geometry parameters are $L_x/H = 9$, $L_y/H = 3.036$ and $W/H = 1.929$. The Reynolds number based on $H$ and the bulk velocity at crest $U_b$ is $Re_b$=5600. The Mach number based on $U_b$ is set to 0.2 to minimize the compressibility. The flow enters the domain from the left side, and the periodic boundaries are connected from right to left. The flow is driven by a uniform body force. In order to match the flow condition with DNS (Direct Numerical Simulation) results [34], the body force is adjusted to acquire the settled mass flow target. When a simulation finished, the body force and mass flow target are also converged.

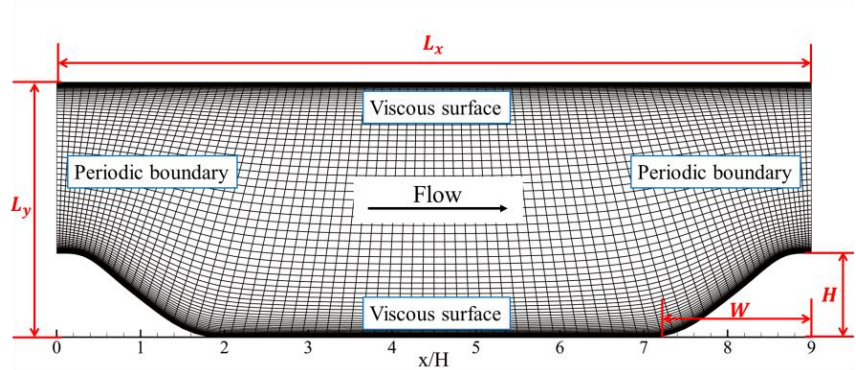

**Fig. 2. Computational domain and boundary conditions**

The flow velocity showed in Fig. 3 are all nodimensionalized by $U_b$. The separation point locates at the top of hill and reattachment point locates at the bottom. The $k - \omega$ model predict a separation bubble much larger than DNS result, which is caused by insufficient mixing from the separating shear layer. In contrast, the two SPF models with turbulent nonequilibrium modification increase the strength of mixing and produce smaller bubbles. Fig. 4 compares the velocity profiles of four turbulent models. The SST and $k - \omega$ models fail to predict the rear part of periodic hill as the velocity profiles show a deviation than the DNS result. While the SPF $k - \overline{v^2} - \omega$ and SPF $k - \omega$ models yield satisfied results. In conclusion, the nonequilibrium modification help produce more turbulent energy in the separating shear layer, which brings forward the reattachment point of the separation bubble and shortens the bubble's length.

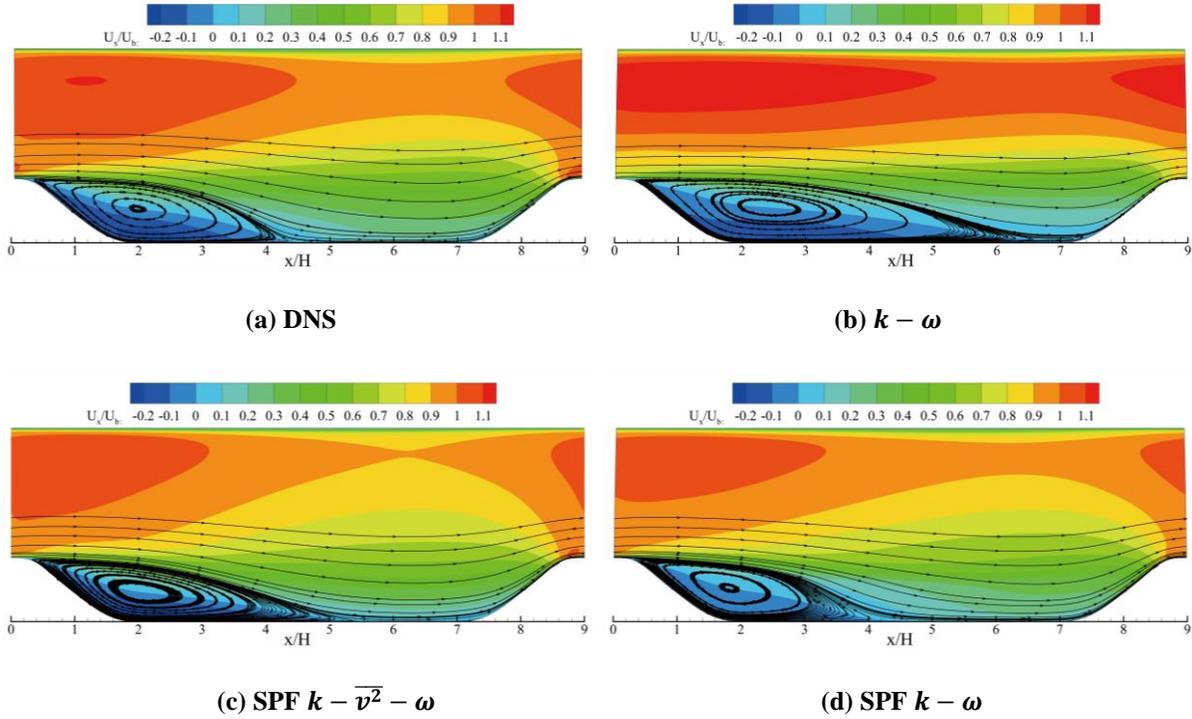

(a) DNS  (b) $k - \omega$

(c) SPF $k - \overline{v^2} - \omega$  (d) SPF $k - \omega$

Fig. 3. Velocity contour and streamlines using different turbulent models

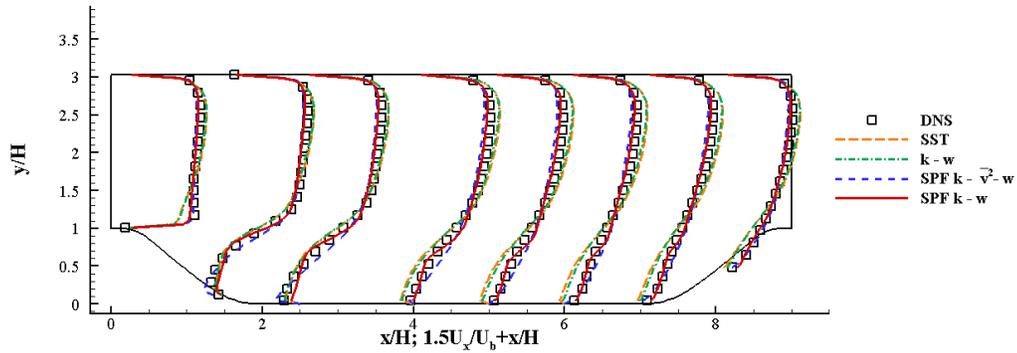

Fig. 4. Velocity profiles along flow direction using different turbulent models

## C. GLC305 Airfoil with 944 Ice

The GLC305 airfoil with ice shape 944 [6] is a typical glaze ice case. The surrounding flow at large angles of attack was considered to have a strong separation. The full-scale ice accretion experiment has been conducted at the NASA John H. Glenn Research Center. Experimental results are gathered from [6] [7] [35]. The surface of ice was considered in a smooth state and the relative ice height is $h_{ice}/c = 3.68\%$. Mesh generation around the complex glaze-ice shape is achieved by solving elliptic equations [36]. The computational grid generated by an in house code in Fig. 5(a) demonstrates satisfactory orthogonality. The current grid for the iced airfoil contains 637 and 133 points over the circumferential and wall-normal directions. Local grid refinement is imposed after the ice tip region to capture the shear layer. The spacing from the first grid layer to the wall is less than $6 \times 10^{-6} c$, which will ensure that $\Delta y^+$ is less than 1.0. The mesh spacing in wall normal direction follow the growth rate of 1.15. Inflow Mach number in this case is 0.12, and Reynolds number based on chord is $3.5 \times 10^6$. Fig. 5(b) illustrates the streamlines at 6° AOA, as predicted with the SPF $k - \omega$ model. The horn upstream of the leading edge shows an obvious critical angle to the incoming flow, which induces a separating shear layer and a large separation bubble, and finally impending stall. The calculated reattachment point at $x/c = 0.48$ is slightly ahead of the measurements where $x/c = 0.53 - 0.60$.

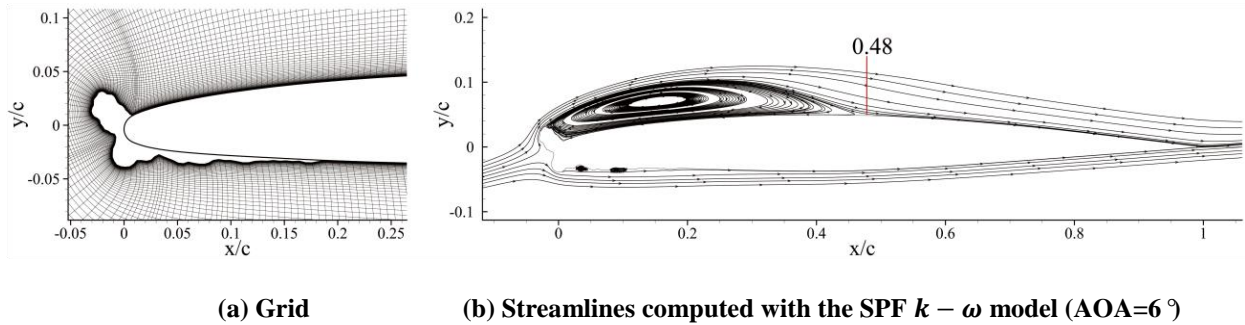

**(a) Grid**　　　　　**(b) Streamlines computed with the SPF $k - \omega$ model (AOA=6°)**

**Fig. 5. Grid and flow field of GLC305 airfoil with horn-shaped ice**

Fig. 6 shows the comparison of the RANS results of five turbulence models SA, SST, $k - \omega$, SPF $k - \overline{v^2} - \omega$ and SPF $k - \omega$. The SA and SST models fail to predict the lift curve because its slope is lower than the experimental data. The stall behaviors at large angles of attack deviate far from the measurements. The $k - \omega$ model performs better than the SST model, but it predicts the inflection point much earlier than the experiment; thus, the maximum $C_L$ is considerably lower than the measured value. Significant improvements in two SPF models are observed. At the AOA of 6 degrees, the relative errors of $C_L$ for the SA [11], SST [12], $k - \omega$ [24], SPF $k - \overline{v^2} - \omega$ and SPF $k - \omega$ models are 25.1%, 36.4%, 17%, 1.9%, and within 1%, respectively. The prediction of the drag coefficient is nearly identical for all models before the stall angle. After the stall point, all four models underpredict the drag coefficient. For moment coefficient prediction, the SST model yields an earlier inflection point, while the other three models capture the critical inflection point at AOA=5 degrees. The two SPF models well predict the decreasing trend of the moment curve after 5 degrees, while the SA and $k - \omega$ models show an early stall and make the moment curve bend at AOA=7 degrees.

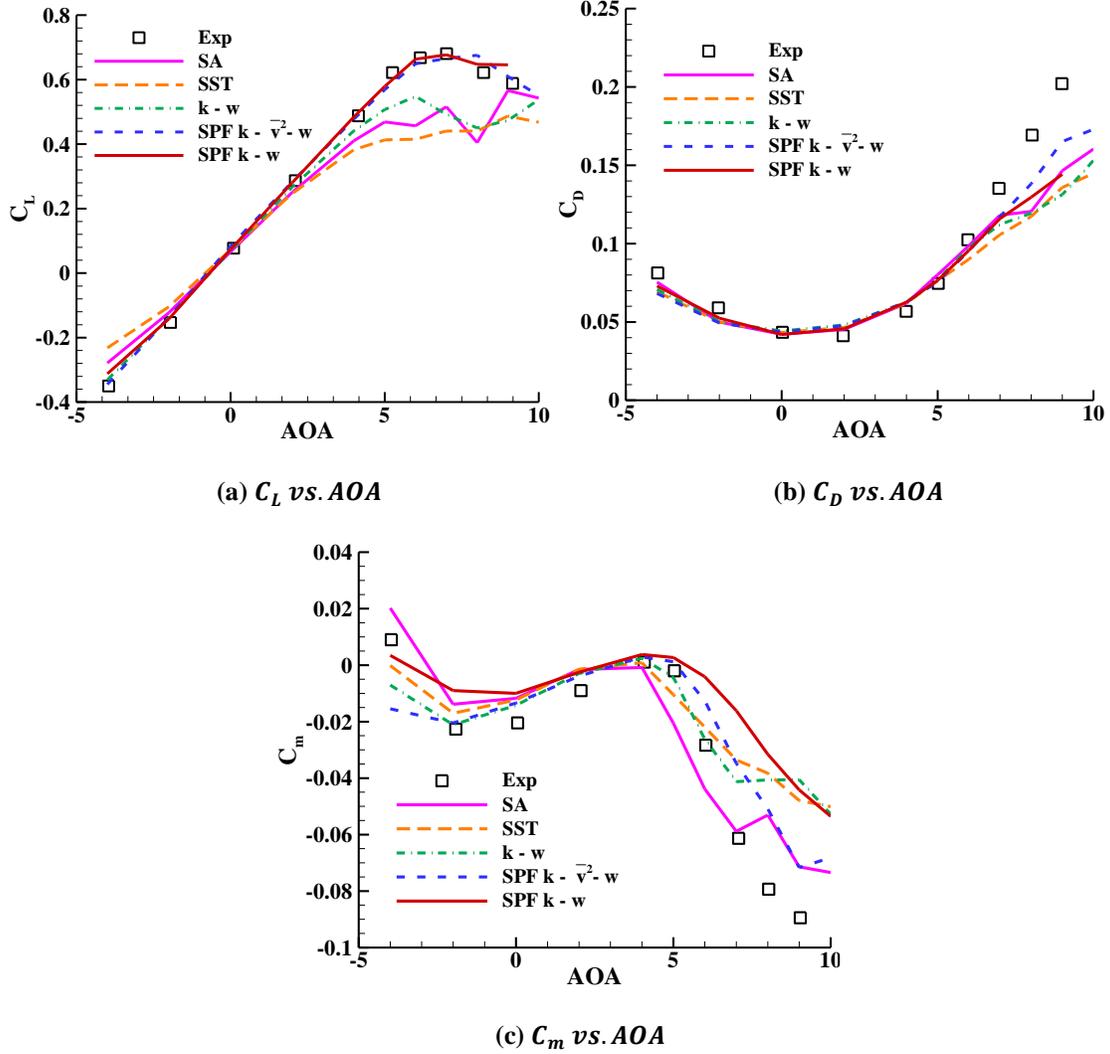

(a) $C_L$ vs. AOA

(b) $C_D$ vs. AOA

(c) $C_m$ vs. AOA

**Fig. 6. Comparisons of turbulent models in evaluating the aerodynamic coefficients of GLC305 with horn-shaped ice (Ma=0.12; Re=$3.5 \times 10^6$)**

Comparing the pressure coefficient (Fig. 7), at AOA=4°, we observe that the short $C_p$ plateau from $x/c \approx -0.02$ to $x/c \approx 0.10$ on upper surface indicates a small separation bubble after the ice horn. At AOA=6°, the length of constant pressure is enlarged with the increasing length of the separation bubble. The SPF $k - \overline{v^2} - \omega$ and SPF $k - \omega$ models have nearly identical satisfying results for the two angles of attack, while the SA, SST and $k - \omega$ models underestimate the height of the suction plateau. At AOA=8°, the airfoil is in a state of full stall, the SA and SST models predict nearly the same results as a persistence low pressure suction platform, which means that separation bubble cover the whole upper surface. The $k - \omega$ model predicts a little bit higher $C_p$ peak, and the distribution of $C_p$ at 8° is nearly the same as $C_p$ at 6°. The SPF $k - \overline{v^2} - \omega$ model shows that the peak pressure decrease from 6° to 8° and the pressure recovery region is enlarged. The SPF $k - \omega$ model still has the largest value of $C_p$ platform, and the pressure distribution at 8° is comparable to 6°.

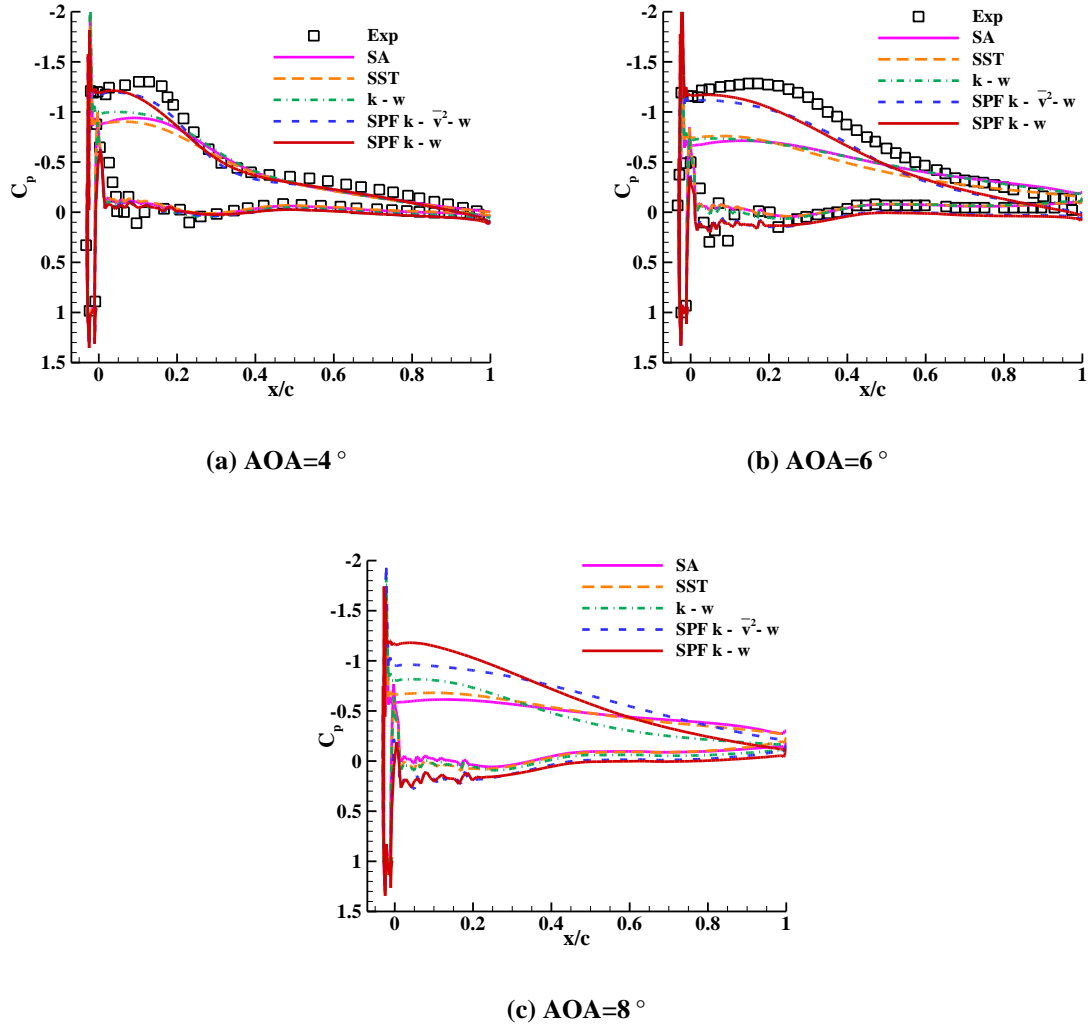

**Fig. 7. Comparisons of turbulent models for the pressure distributions of GLC305 with horn-shaped ice**

Fig. 8 illustrates the contour of normalized turbulence intensity. Fig. 9 shows the profiles of turbulence intensity at $x/c = 0.1, 0.35$ and $0.75$. These figures illustrate how the turbulent intensity evolves through the upper surface of the iced airfoil. According to the experimental observation, the flow separates at the ice tip. The shear layer breakup and transition process occurring in the streamwise flow gradually produces more $U_{rms}$, which enhances the mixing until reattachment. When $x/c \leq 0.1$, the proposed shear stress limiter of the two SPF models perform like that of $k - \omega$ model and acquire good profiles compared to experimental data. In the separated shear layer area ($0.1 < x/c < 0.6$), the nonequilibrium of turbulence is dominant, and the modified term $f_{NE}$ of SPF models is active to produce more turbulence intensity. The profiles of turbulence intensity at $x/c = 0.35$ and $0.75$ show that the modified turbulence models well predict the mixing process in the separating shear layer. Although the SPF $k - \omega$ model does not describe the transition phenomena, it yields a satisfied development of the turbulence intensity and a peak value of $U_{rms}/U_{inf}$ similar to the transitional model SPF $k - \overline{v^2} - \omega$. The predicted peak value of $U_{rms}/U_{inf}$ by $k - \omega$ is much lower than the experimental results.

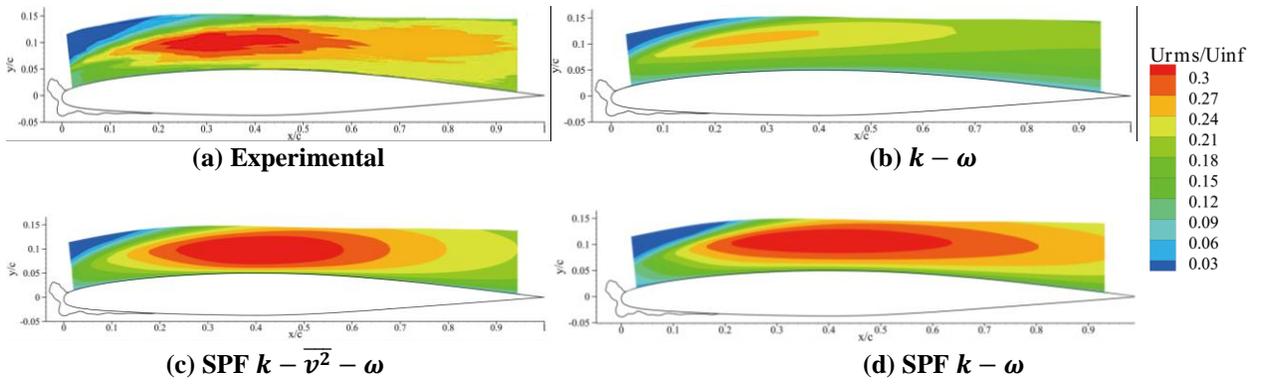

(a) Experimental  (b) $k - \omega$

(c) SPF $k - \overline{v^2} - \omega$  (d) SPF $k - \omega$

Fig. 8. The contour of normalized time-averaged turbulence intensity at AOA=6 °

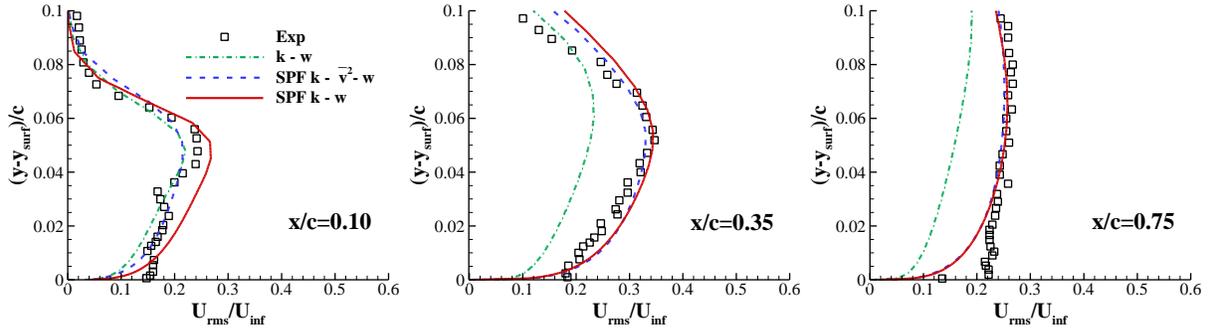

Fig. 9. Turbulence intensity profiles at AOA=6 °

The nonequilibrium characteristic of the turbulence is evaluated by the production-to-dissipation ratio of turbulent kinetic energy. The ratio that judge the nonequilibrium region in the context of the $k - \omega$ model framework is obtained as Eq. (10), while the ratio of the $k - \overline{v^2} - \omega$ model is obtained as Eq. (5). In the separating shear layer region, the production of $k$ should be significantly larger than $\varepsilon$. But the ratio of $P_k/\varepsilon$ in homogenous shear flow region is around $1.6 \pm 0.2$. Fig. 10 illustrates the evaluated production-to-dissipation ratio using the $k - \omega$, SPF $k - \overline{v^2} - \omega$ and SPF $k-\omega$ models. The $k-\omega$ model predicts the ratio of production-to-dissipation smaller than 2.0 in the separating shear layer, which indicates equilibrium turbulence. In contrast, the SPF $k-\overline{v^2}-\omega$ and SPF $k-\omega$ models predict the ratio of $P_k/\varepsilon$ larger than 3.0 in the separating shear layer where the flow is in the nonequilibrium state. Further analysis shows that $k-\omega$ was mainly calibrated in the cases of homogenous shear flow; consequently, this model lacks the corresponding mechanism to simulate the nonequilibrium phenomenon. With the modification terms of two SPF models acting on the separating shear layer, the nonequilibrium characteristic of turbulence can be predicted.

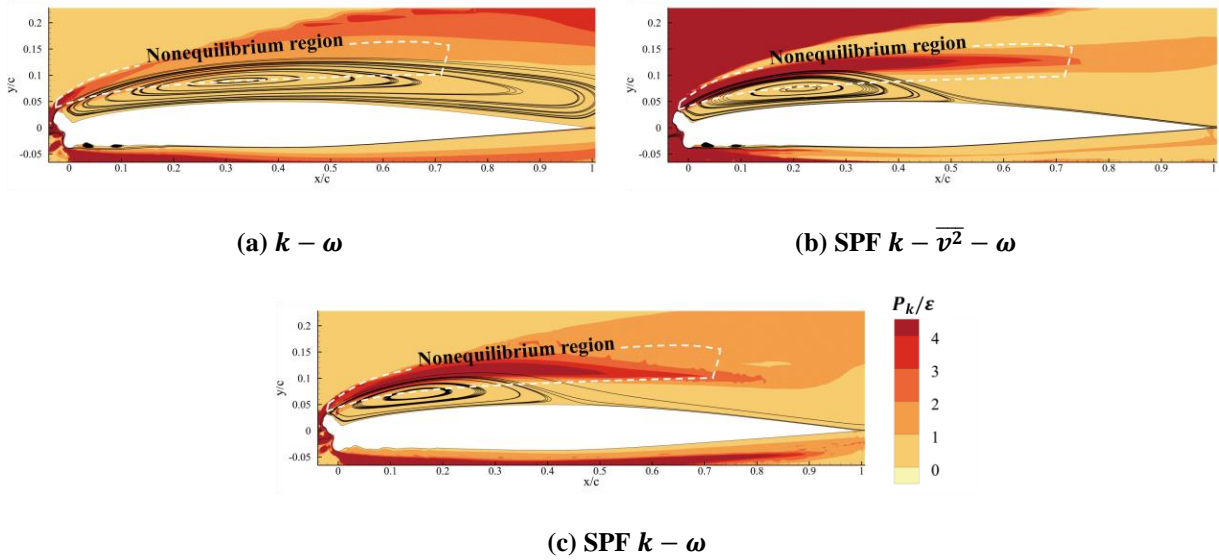

(a) $k-\omega$

(b) SPF $k-\overline{v^2}-\omega$

(c) SPF $k-\omega$

Fig. 10. Contours of $P_k/\varepsilon$ at AOA=6 ° as obtained using three turbulent models

Fig. 11 presents the grid convergence performance of the SPF $k-\omega$ model. The three levels of grid densities are 361×69, 457×97, and 637×133, respectively. A nearly identical slope of lifting curves for the three grid levels is observed. In addition, the fine grid better evaluates the stall performance after 8 degrees because it shows a decreasing trend of $C_L$. The coarse grid over-predicts $C_L$ at negative AOAs because the separated region of the lower surface is not well predicted. From Fig. 11(b), the predicted drag decreases from the coarse grid to the fine grid. The medium and fine grids could yield drag values that are in good agreement with the measurement. For the pressure coefficient at 4 degrees, the coarse and medium grids well predict the height of the pressure platform while the fine grid yields a satisfying pressure recovery. In conclusion, the proposed SPF $k-\omega$ model exhibits a grid convergence performance.

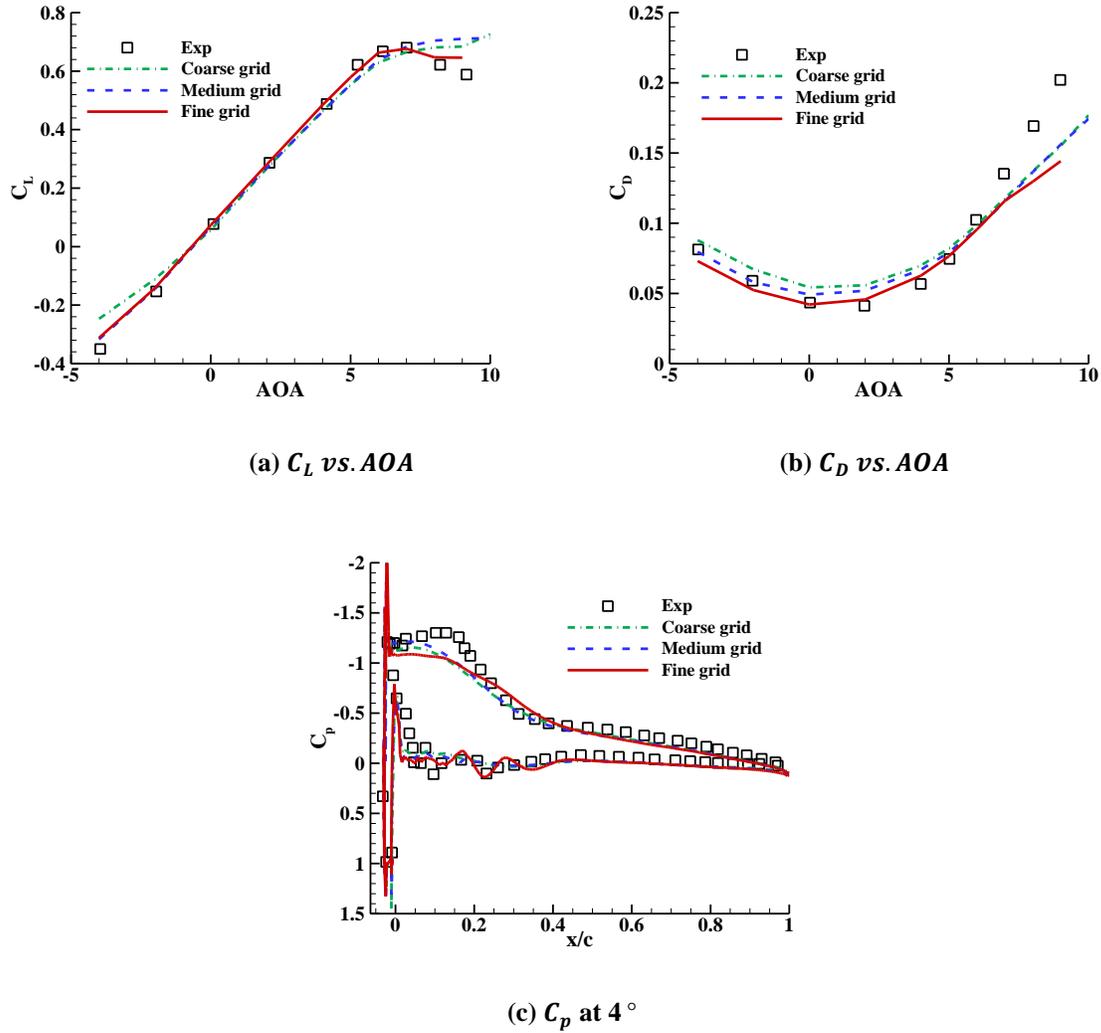

**(a)** $C_L$ vs. AOA

**(b)** $C_D$ vs. AOA

**(c)** $C_p$ at 4°

**Fig. 11.   Grid convergence study for GLC305 with horn-shaped ice using the SPF $k-\omega$ model**

### D. NACA0012 Airfoil with Simulated Glaze Ice Accretion

The second iced airfoil case is an NACA0012 airfoil with simulated ice accretion. Two-dimensional experimental results were taken by Bragg et al. [4]. The ice accretion was a simulation of that measured in the NASA Icing Research Tunnel. The experiment icing conditions were: free-stream velocity of 58 m/s, AOA of 4 degrees, ice accretion time of 5 minutes, volume median diameter droplet of 20 microns, Liquid Water Content (LWC)=2.1 g/m$^3$, and temperature of 18°F. Under these conditions, the ice is considered glaze and can significantly alter the airfoil flow field. For aerodynamic measurements, the incoming Mach number in the wind tunnel is 0.12, and the Reynolds number is $1.5 \times 10^6$. Fig. 12(a) shows the grid of this iced airfoil. The grid contains 457 and 97 points over the circumferential and wall-normal directions, respectively. From Fig. 12(b), the separation point locates at the ice tip; then, a stable separation bubble is calculated. The reattachment point of separation flow is at $x/c = 0.3$.

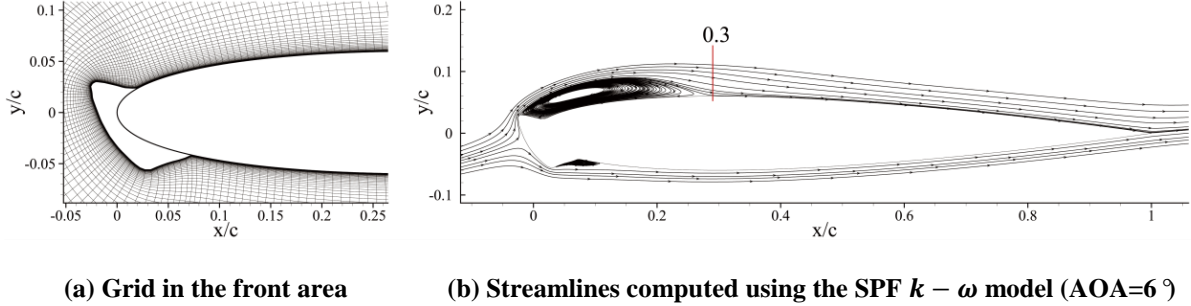

(a) Grid in the front area  (b) Streamlines computed using the SPF $k - \omega$ model (AOA=6°)

Fig. 12. Grid and flow field of NACA0012 airfoil with simulated glaze ice

Fig. 13(a) depicts the lift curve in a large range of angles of attack. The SA and SST models predict the stall angle earlier before the experimental results. The $k - \omega$, SPF $k - \omega$ and SPF $k - \overline{v^2} - \omega$ models well predict the stall angle of attack at AOA=9 degrees. However, the $k - \omega$ model fails in predicting the maximum lift coefficient. For $C_L$ at negative angles, SPF $k - \overline{v^2} - \omega$ shows the best result compared with the measurements. The relative error of the maximum lift coefficients of the SA, SST and $k - \omega$ model are 35.4%, 39.93% and 19.6%, while the errors for SPF $k - \overline{v^2} - \omega$ and SPF $k - \omega$ are 7.1% and 12.5%, respectively. Fig. 13(b) shows the drag coefficient prediction. The $k - \omega$ model shows a large deviation between the predicted results and the measurements. The SPF $k - \overline{v^2} - \omega$ provides the best result, since the drag polar curve is consistent with the experimental data overall. The SPF $k - \omega$ model over-predicts the drag at negative $C_L$ but it well predicts the value at positive $C_L$.

Fig. 13(c) shows the comparison of the pressure coefficient. From the experimental observation, a region of almost constant pressure on the upper surface extends from the leading edge to $x/c = 0.08$, which indicates a separation bubble after the leading-edge. Although the bubble is quite large, it acts as a classical airfoil laminar separation bubble. The shear layer is initially laminar with a transition process after the ice tip, and the constant pressure indicates the process [28]. The turbulent mixing in the separating shear layer leads to reattachment downstream after some degree of pressure recovery. All five models can capture the constant-pressure region. The transitional SPF $k - \overline{v^2} - \omega$ model produces a longer constant-pressure region than the other models.

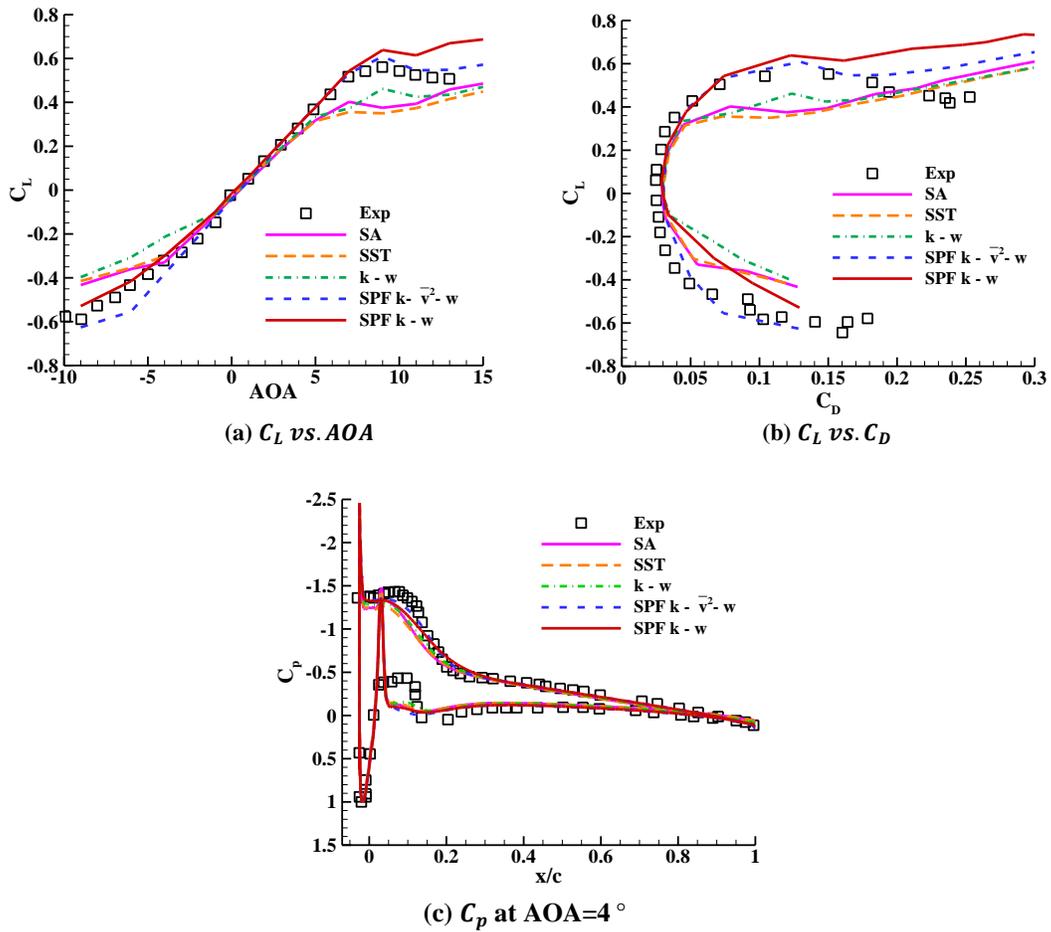

**Fig. 13.** Comparisons of turbulent models in evaluating the aerodynamic coefficients of NACA0012 with simulated ice (Ma=0.12; Re=$1.5 \times 10^6$)

Fig. 14 shows the grid convergence performance of the SPF $k-\omega$ model. The three levels of grid densities are 457×97, 637×133, and 1037×225 respectively. The main tendency of the relationship between the grid density and lift coefficient can be obtained. When increasing the number of girds, the value of maximum lift will increase. But the stall angles are constant. The prediction of drag value improves a lot with the increase of gird nodes. Especially at the negative angle of attacks.

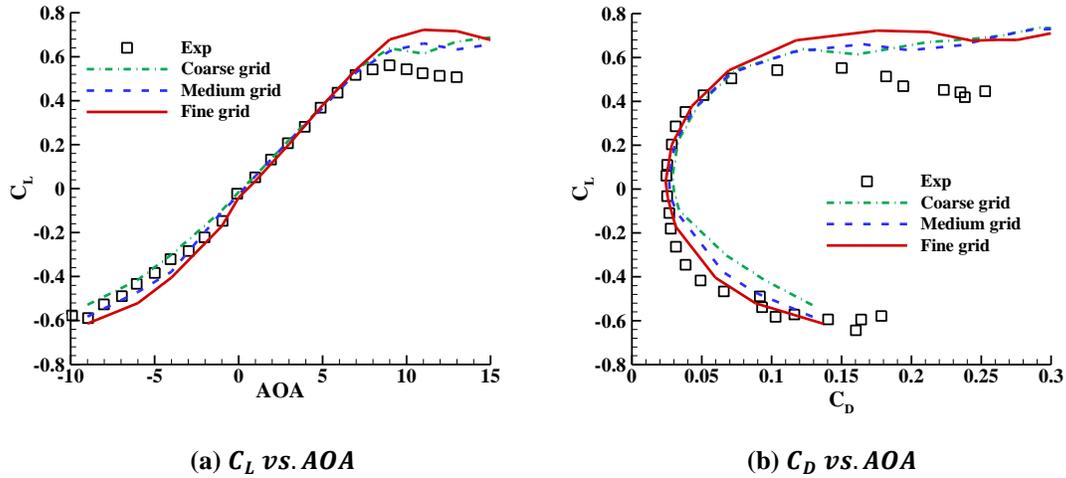

(a) $C_L\ vs.\ AOA$  (b) $C_D\ vs.\ AOA$

Fig. 14.  Grid convergence study for NACA0012 with simulated ice using the SPF $k-\omega$ model

### E. Ice Accretion on a Swept Wing

The three-dimensional iced wing is a semispan wing with a chord of 0.381 m and a span of 0.8936 m. The NACA0012 airfoil with simulated glaze ice of the last section is used on this 30-degree swept wing. The iced airfoil section is in the plane perpendicular to the leading edge. The considered wing geometry has an untwisted, untapered planform shape. Experimental results for this case are produced by Bragg et al. [1] [8] [9]. Fig. 15 shows the computational domain and boundary conditions. The size of the domain is approximately equivalent to the wind tunnel. The inflow/outflow boundary is set at the front and back faces (colored purple). The model in wind tunnel is a semispan wing with a splitter wall at the root location, and a boundary suction system is employed at the front. The splitter wall is applied to reduce the impact of the boundary layer of the tunnel wall. But the intersection of the wing and tunnel wall still results in the development of a vortical flow reducing the effective angle of attack at the wing root. To properly capture this phenomenon in this region, a symmetry plane condition is applied at the front, and a viscous wall condition (colored blue) is implemented at the wing root-splitter plate junction.

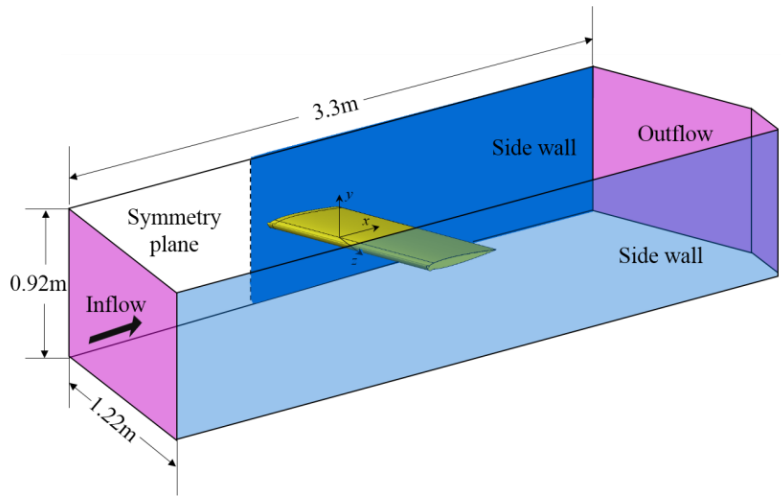

**Fig. 15.** Computational domain and boundary conditional of the ice accretion on the NACA0012 swept wing

The three-dimensional grid in Fig. 16 adopts a two-dimensional C-grid in a region not far away from the wing surface. The grid generation routines perform the necessary interpolation from 2D iced airfoil to enrich the input wing shape. Mesh of the outer field is filled by H-grid. The spanwise spacing of the nodal plane in the region of the wing root is notably small to capture the observed sidewall effect in the wind tunnel. The number of grid cells is approximately 3.5 million. The inflow velocity operates at speeds of Mach number of 0.12. The tested Reynolds number is $4.92 \times 10^6$ per meter. The turbulence intensity of inflow is set as 0.05% according to the experiment.

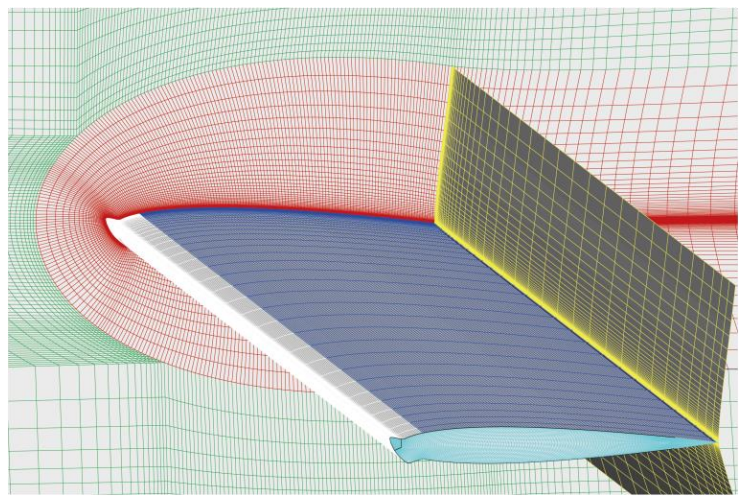

**Fig. 16.** Schematic of the grid for the iced swept wing

Fig. 17 shows the prediction of the integrated aerodynamic coefficients using three turbulent models. The experimental data for the lift coefficient continue increasing with the increase in AOA, which does not show an obvious stall onset point. The models except $k-\omega$ have identical trends. The SA and SST have better performance compared with 2D sectional Case D. However, the $k-\omega$ model produces an inflection point at AOA=12 degrees and a lift loss at larger angles. All five models predict excessive drag values compared to the experimental data. The drag polar curve of the SPF $k-\overline{v^2}-\omega$ model is closer to the experimental data, and the deviation of $C_D$ from measurements is maintained at a stable value over the range of $C_L$.

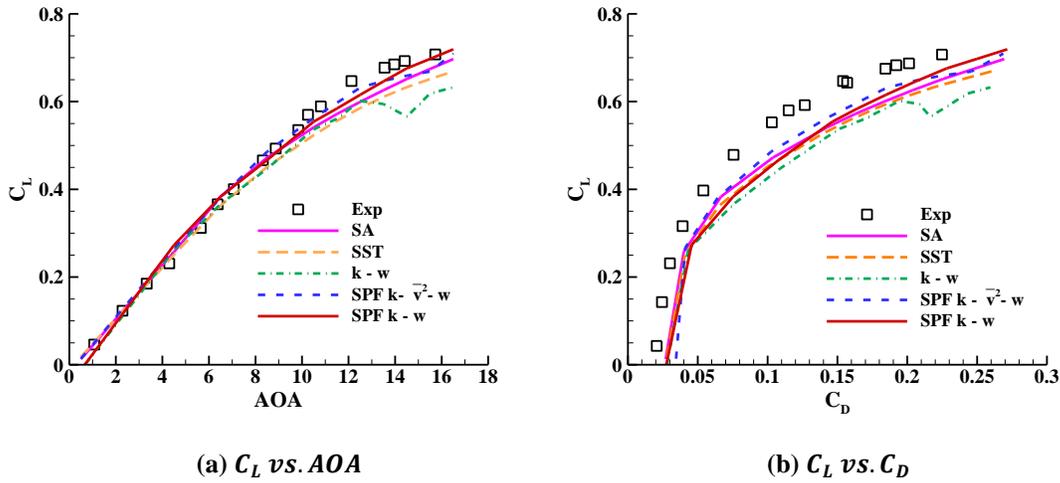

(a) $C_L\ vs.AOA$          (b) $C_L\ vs.C_D$

**Fig. 17. Comparisons of turbulent models in evaluating the aerodynamic coefficients of the iced swept wing (Ma=0.12, Re=$4.92\times 10^6$/m)**

The model pressures were converted to pressure coefficients using the measured tunnel dynamic pressure taken from each scanivalve. The pressure tabs of measurements are located in 5 rows [1], as shown in Fig. 18. The pressure coefficients were integrated over the surface of the model to produce the section lift coefficient. From Fig. 19, the calculated and measured spanwise load distributions are notably consistent with each other using the two separation fixed models. The loads gradually reduce from the root to tip sections because of the swept effect and wingtip vortex effect. The SPF $k-\omega$ model shows a similar load distribution with the $k-\omega$ model at the AOA of 4 degrees. However, at a large angle of attack of 8 degrees, the SPF $k-\omega$ model greatly improves from the $k-\omega$ model because the prediction of load at the outboard sections is closer to the experimental data.

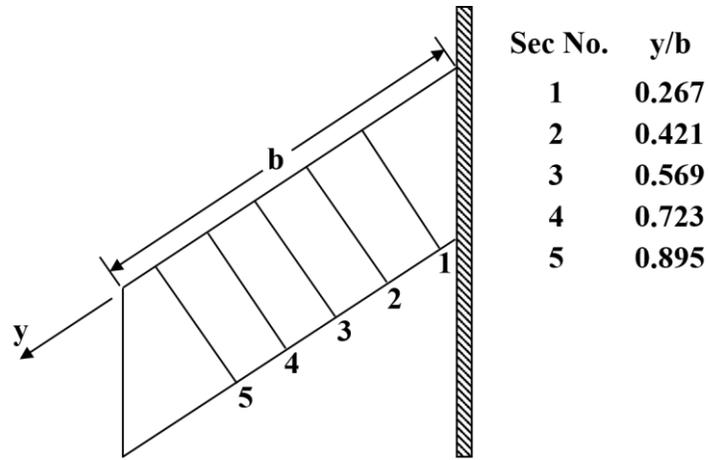

**Fig. 18.** Surface pressure taps installed on the experimental geometry

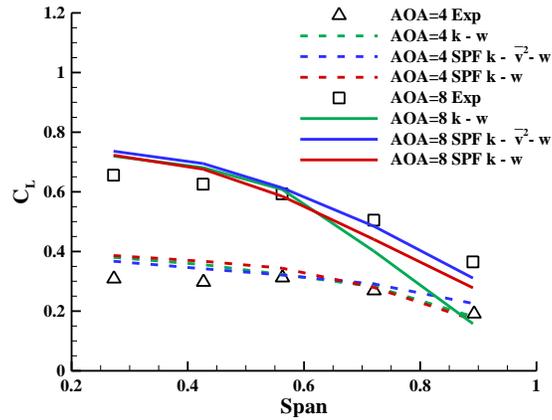

**Fig. 19.** Comparison of different turbulent models for evaluating the spanwise load distributions at AOA = 4 and 8 degrees

The flow visualization of the 30-degree swept wing illustrates a very three-dimensional flow field. Experimental and computational surface flow visualizations are shown in Fig. 20. The computational results are postprocessed from the simulated oil flow by surface streamlines introduced into the RANS flow field. The main feature of the flow at AOA of 8 degrees is the leading-edge vortex. The leading-edge vortex grows in size when it moves from the root to the tip. Experimental and computational results show that the interaction of the large leading-edge vortex and tip vortex causes a complex flow at the wingtip. The flow first moves forward towards the leading edge and then returns towards the trailing edge. The SPF $k - \overline{v^2} - \omega$ and SPF $k - \omega$ models preciously capture the above phenomenon. However, the $k - \omega$ model does not show apparent turning back streamlines in the stall region (inside the red dotted box), which indicates that the separation bubble at the wingtip is overpredicted by the $k - \omega$ model. The off-body streamlines computed by the SPF $k - \overline{v^2} - \omega$ model in Fig. 21 show that the center of the vortical flow moves back along the chord as it progresses from the root to the tip. At the AOA of 4 degrees, the flow separates at the tip of ice and reattaches at nearly 10% chord. At the AOA of 8 degrees, the evolving process of the 3D vortex influenced by the swept effects is presented, which is consistent with the experimental observation. The SPF $k - \overline{v^2} - \omega$ and SPF $k - \omega$ models computes nearly the same large vortex developed from the leading edge. However, the $k - \omega$ model predicts a vortex that is highly disturbed by the cross flow, which cause the discontinues pressure on the upper surface.

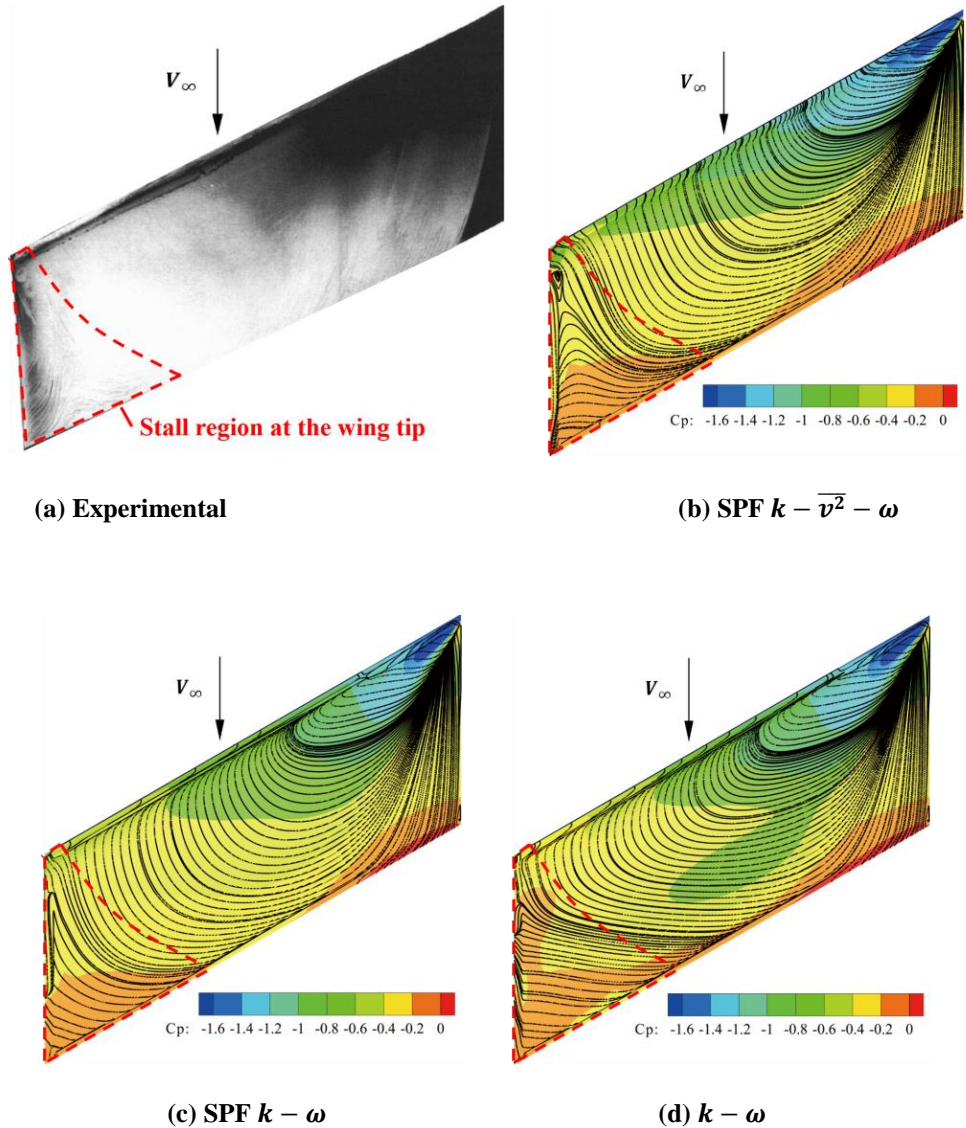

(a) Experimental        (b) SPF $k - \overline{v^2} - \omega$

(c) SPF $k - \omega$        (d) $k - \omega$

**Fig. 20.    Surface flow visualization on the iced swept wing at AOA=8 °**

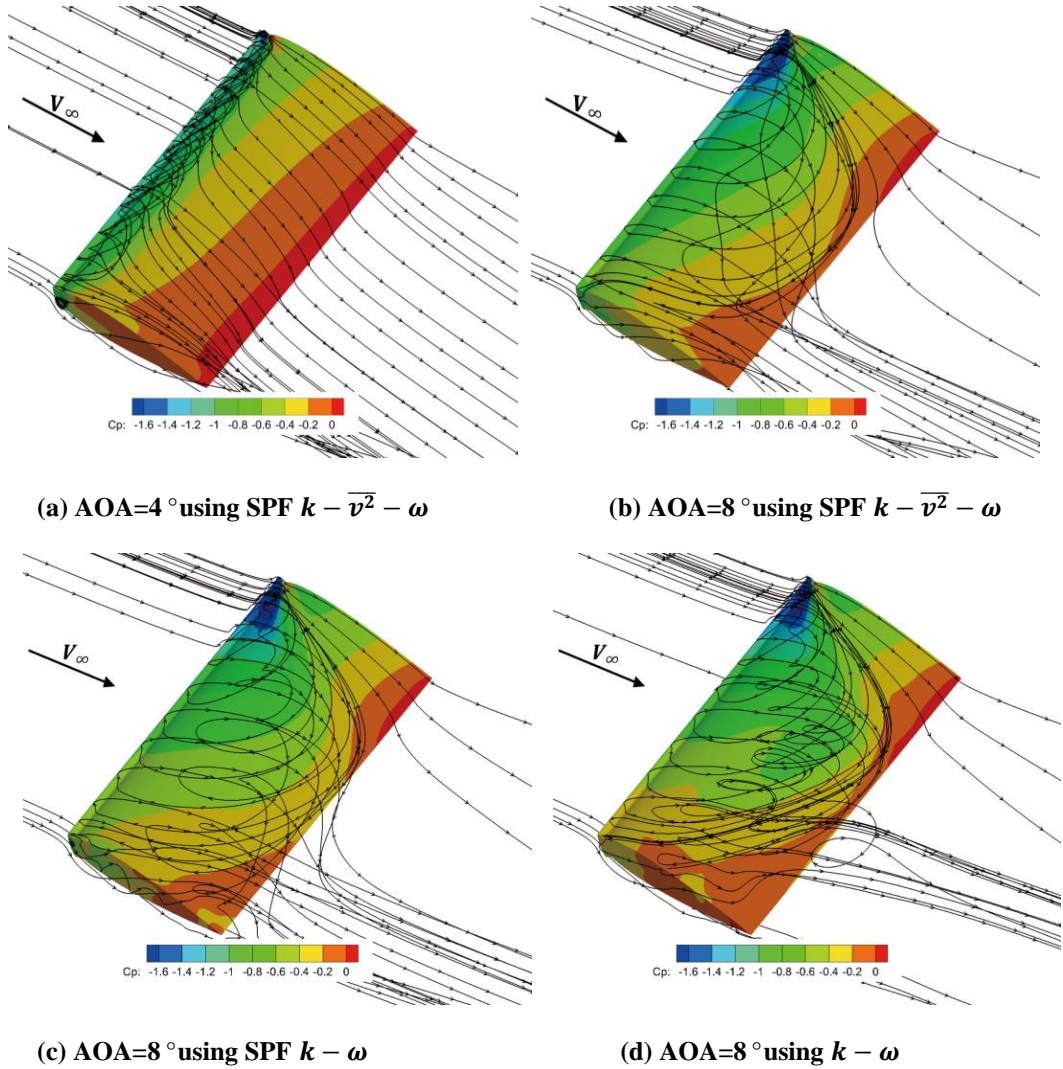

(a) AOA=4 °using SPF $k - \overline{v^2} - \omega$    (b) AOA=8 °using SPF $k - \overline{v^2} - \omega$

(c) AOA=8 °using SPF $k - \omega$    (d) AOA=8 °using $k - \omega$

**Fig. 21.    Off-body particle traces for the iced swept wing using different turbulent models**

The pressure distributions at several spanwise locations for the angle of attack of 4 and 8 degrees are shown in Fig. 22. The experimental and computational results show a flat region of outboard $C_p$, which indicates that the swept model first stalls at the wing tip. At AOA=4 degrees, the results of SPF $k - \overline{v^2} - \omega$ are consistent with experimental data at all span locations. The suction peak values are well captured. The $k - \omega$ and SPF $k - \omega$ models can produce satisfied pressure distributions at the inboard wing but fail in the outboard wing because the spanwise separation bubble is overpredicted. At AOA=8 degrees, the SPF $k - \overline{v^2} - \omega$ yields good pressure peaks near the leading edge except for the section at 27% span possibly because the flow is not fully accelerated by the disturbance of the sidewall. $k - \omega$ and SPF $k - \omega$ have similar results at the first two spanwise locations, but the $k - \omega$ model shows a fluctuation of $C_p$ near the trailing edge region at the other three spanwise locations. Thus, SPF $k - \overline{v^2} - \omega$ shows the best performance in the calculation of the iced wing and can produce an accurate pressure distribution. The SPF $k - \omega$ model yields satisfactory integrated aerodynamic coefficients because it delays the stall at the wingtip, but it does not predict a sufficient suction peak of pressure at the leading edge. The $k - \omega$ model without separation fix fails in predicting the stall performance of the iced wing.

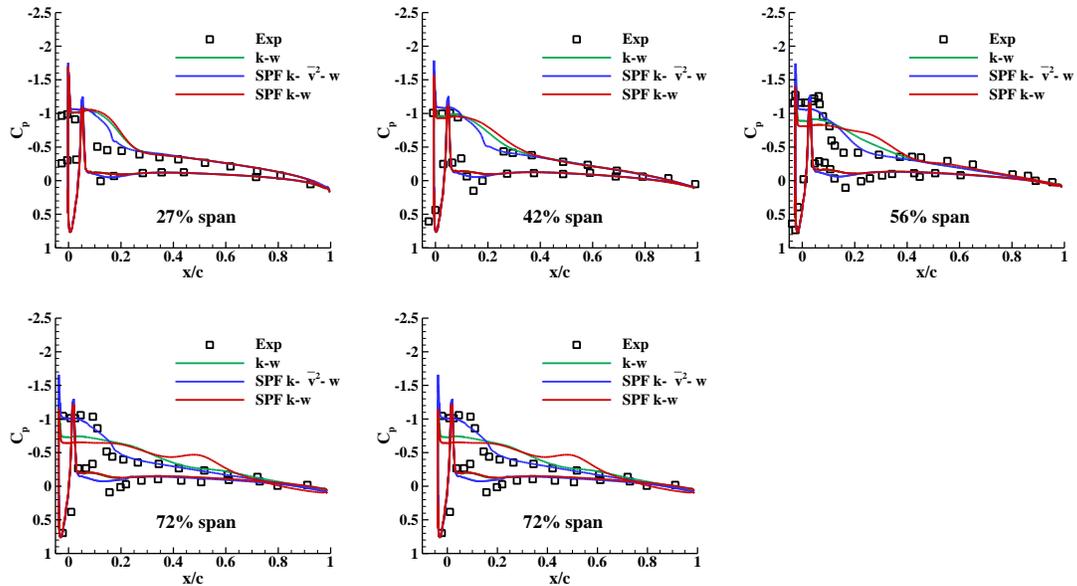

(a) AOA=4°

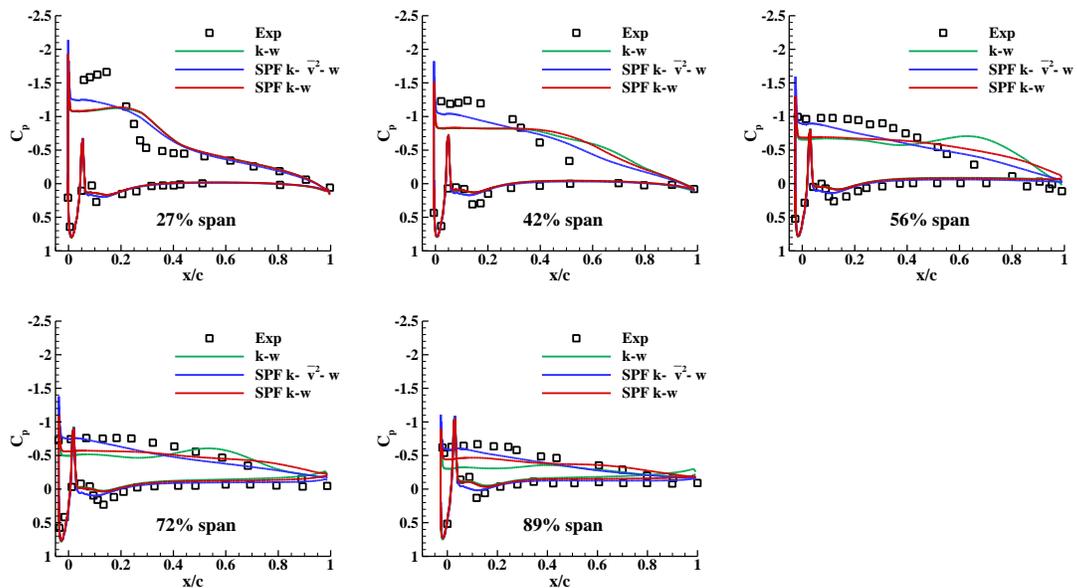

(b) AOA=8°

**Fig. 22.    Surface pressure distribution for the iced swept wing using different turbulent models**

## V. Conclusions

Aerodynamic prediction of iced airfoil and wing has been done by proposed linear eddy viscosity RANS models. Some ad hoc separation fix terms of turbulence models are proposed with a focus on the nonequilibrium characteristic of turbulence in a separating shear layer. The nonequilibrium effect refers to a state where the turbulent kinetic energy production is much larger than its dissipation. The baseline $k-\omega$ and $k-\overline{v^2}-\omega$ models exhibit poor performance in simulating the nonequilibrium characteristic of turbulence and under-predict the shear stress in the separating region of full turbulence. The effects of the nonequilibrium characteristic are modeled by boosting the destruction term of the ω-equation in the recognized separating shear layer region. Shear stress limiters are adopted to appropriately simulate the beginning transition process of the shear layer when turbulence is under development. Implementing these fixed terms can improve the ability of the original models to predict the separating shear layer and will not deteriorate the performance of the models in predicting the log-layer or basic free shear flows. The developed separation fixed terms for two baseline turbulence models have similar concise forms and can be easily added to an existing CFD program.

The glaze ice accretion in this paper severely affects aerodynamics performances. The proposed SPF $k-\overline{v^2}-\omega$ and SPF $k-\omega$ models could produce satisfying results for aerodynamic coefficients. And some key features of the averaged flow field of iced airfoils, such as the separation bubble and pressure distribution can be well simulated. However, the $k-\omega$ model exhibits some faults in predicting the stall performance of the cases.

The observed swept iced wing has a complex 3D flow field. Significant cross-flow is observed in the vortex core. The SPF $k-\overline{v^2}-\omega$ and SPF $k-\omega$ models predict satisfying integrated aerodynamic coefficients and pressure distributions. Flow visualization results further illustrate that the modified RANS turbulence models well predict the separation region of the outboard wing compared to the original $k-\omega$ model.

## Acknowledgments

This work is supported by the National Natural Science Foundation of China under Grant Nos. 11872230 and 91852108.